\documentclass[final]{aipproc}
\usepackage{graphicx}
\usepackage{epstopdf}

\layoutstyle{8x11single}

\begin{document}

\title{Solution of three-dimensional Faddeev equations: ultracold Helium trimer calculations with a public
quantum three-body code}

\classification{21.45.-v, 31.15.ac, 34.10.+x, 36.90.+f, 02.70.Jn}
\keywords{three-body atomic systems, Faddeev equations, Helium trimer, ultracold collisions}

\author{E.~A. Kolganova}{
  address={BLTP JINR, 141980 Dubna, Moscow region, Russia}
}

\author{V. Roudnev}{
  address={Department of Physics and Astronomy, University of Kentucky, Lexington,
Kentucky, 40506-0055, USA} }

\author{M. Cavagnero}{
  address={Department of Physics and Astronomy, University of Kentucky, Lexington,
Kentucky, 40506-0055, USA}
 % ,altaddress={<author1 address>} % additional visiting address
}
\begin{abstract}
We present an illustration of using a quantum three-body code being prepared for public
release. The code is based on iterative solving of the three-dimensional Faddeev
equations. The code is easy to use and allows users to perform highly-accurate
calculations of quantum three-body systems. The previously known results for He$_3$ ground
state are well reproduced by the code.
\end{abstract}

\maketitle

%%%%%%%%%%%%%%%%%%%%%%%%%%%%%%%%%%%%%%%%%%%%
%% MAINMATTER
%%%%%%%%%%%%%%%%%%%%%%%%%%%%%%%%%%%%%%%%%%%%

\section{Introduction}

The quantum few-body problem is important for investigating physical processes at
practically all possible length and energy scales. For instance, three-body models can be
employed for describing nuclear reactions~\cite{Friar,dd,alphaBe,Lambda}, electron- and
positron-atom collisions~\cite{Hu,Rescigno}, and chemical reactions~\cite{HNe2}. The
developments of the last decade demonstrated the importance of three-body processes for
understanding the dynamics of ultra-cold gases~\cite{EsryGreen}. An ability to solve a
few-body problem directly would also be beneficial to theorists for testing, for
instance, effective-field theories~\cite{Braaten,Phillips}.

The three-body problem, however, has sufficient intrinsic complexity that it often
inhibits or prevents non-experts in few-body calculations from considering realistic
three-body models and from employing physically correct representations~\cite{BadHelium}.
Accordingly, a standard, easily operable and rigorously constructed tool for three-body
calculations will be beneficial for a broad physical community. Such a tool
should be tested independently to ensure its usability and applicability.
This work is a result of a collaboration between the authors of this tool, developed at
the University of Kentucky, and a research group in JINR performing independent tests.

In the following sections we describe the equations being solved and report the results
of the tests we have performed.

\section{Formalism}
The three-body code being tested is based on solving Faddeev equations in configuration
space. The complete and mathematically rigorous theory of Faddeev equations can be found
in books on the topic~\cite{MerkFadd,Integral,SchmidtZigelman}. Here we only sketch out
the gross features important for understanding and using the code.

We start by describing the physical model from the three-body Hamiltonian
\begin{equation}
\label{eq:Hamiltonian}
        H=H_0+V_{\rm 3b}({\bf x}_i,{\bf y}_i )+\sum _i V_i ({\bf x}_i )\,,
\end{equation}
where $H_0$ stands for the kinetic energy of the three particles, $V_i ({\bf x}_i )$ is
the interaction potential acting in the pair $i$, and $V_{\rm 3b}({\bf x}_i,{\bf y}_i )$
is a short-range three-body interaction. (In the following description the latter will be
omitted only for simplicity. Taking into account the three-body interactions, however,
does not produce any practical or principal difficulties.) The configuration space of the
three particles is described in terms of 3 sets of Jacobi coordinates
\begin{equation}
\begin{array}{rcl}\cr
           {\bf x}_i&=&\displaystyle\left(\frac{2m_j m_k }
                {m_j+m_k}\right)^{1/2}
                ({\bf r}_j-{\bf r}_k)\cr
        {\bf y}_i&=&\displaystyle\left(\frac{2m_i (m_j +m_k )}
             {m_i +m_j +m_k }\right )^{1/2}
                \left({\bf r}_i-\displaystyle
               \frac{m_j {\bf r}_j+m_k
                {\bf r}_k}{m_j+m_k }\right)
\end{array}
\label{Jacoord}
\end{equation}
The set of coordinates $i$ describes a partitioning of the three particles into a pair
$(jk)$ and a separate particle $i$. Faddeev decomposition represents the wave function
$\Psi$ in terms of a sum over all possible partitioning of the three-body system
\begin{equation}
  \Psi=\sum_i  \Phi_i({\bf x}_i, {\bf y}_i) \, .
  \label{eq:WF}
\end{equation}
Faddeev components, $\Phi_i$, satisfy the following set of equations~\cite{MerkFadd}
\begin{equation}
(-\Delta_{\bf x} -\Delta_{\bf y}+ V_i({\bf x}_i) - E) \Phi_i({\bf x}_i, {\bf y}_i) =
  V_i({\bf x}_i) \sum_{k\ne i}  \Phi_k({\bf x}_k, {\bf y}_k) \ ,
\label{eq:Faddeev}
\end{equation}
where ${\bf x}_i$ and ${\bf y}_i$ are mass-weighted Jacobi coordinates, $V_i$ is the
interaction potential in the $i$-th pair and $E$ is the total energy of the system. It is
not difficult to prove that the exact wave function of the three body system can be
uniquely constructed from the Faddeev components by means of Eq.~(\ref{eq:WF}).

The equations in six-dimensional space can hardly be solved directly, and some partial analysis
is necessary. We consider the states with zero total angular momentum.
The angular degrees of freedom corresponding to  collective rotation of the three-body system
can be separated \cite{Kvits} and the kinetic energy operator reduces to
\begin{equation}
\label{H00}
        H_{0}=-\frac{\partial ^{2}}
        {\partial x^{2}}-\frac{\partial ^{2}}
        {\partial y^{2}}-(\frac{1}{x^{2}}
        +\frac{1}{y^{2}})\frac{\partial }
        {\partial z}(1-z^{2})^{\frac{1}{2}}
        \frac{\partial }{\partial z}\,,
\end{equation}
where $x$, $y$ and $z$ are so called intrinsic coordinates
\begin{equation}
\label{IntrCoord}
        x=|{\bf x}|,\quad y=|{\bf y}|,\quad
                z=\frac{({\bf x},{\bf y})}{xy},\qquad
                x,y\in [0,\infty ),\ \  z\in [-1,1] \ \ .
\end{equation}
In the case of identical bosons Faddeev components take identical functional form, which makes it
possible to reduce the system of three equations (\ref{eq:Faddeev}) to one equation
\begin{equation}
\label{EqFaddTAM1} \displaystyle
   (H_0+V(x)-E)\phi(x,y,z)
   =
   -V(x) P \phi(x,y,z) \,,
\end{equation}
where
\[P \phi(x,y,z)
      \equiv
      xy(
        \frac{\phi(x^+,y^+,z^+)}{x^+ y^+}
        +\frac{\phi(x^-,y^-,z^-))}{x^- y^-}
        )
\]
and $x^\pm(x,y,z)$, $y^\pm(x,y,z)$, and $z^\pm(x,y,z)$ are
\begin{equation}
\label{xyzpm}
  \begin{array}{rcl} \displaystyle
     x^\pm (x,y,z)& = & \displaystyle
                        (\frac{1}{4}x^2
                              +
                         \frac{3}{4}y^2
                              \mp
                         \frac{\sqrt{3}}{2}xyz
                         )^{1/2}
                         \; ,    \\
     y^\pm (x,y,z)& = & \displaystyle
                        (\frac{3}{4}x^2
                              +
                         \frac{1}{4}y^2
                               \pm
                         \frac{\sqrt{3}}{2}xyz
                         )^{1/2}
                         \; ,   \\
     z^\pm (x,y,z)& = & \displaystyle
                        \frac{\displaystyle\pm \frac{\sqrt{3}}{4}x^2
                              \mp \frac{\sqrt{3}}{4}y^2
                               - \displaystyle \frac{1}{2}xyz}
                             {x^{\pm }(x,y,z)\, y^{\pm }(x,y,z)}\; .
  \end{array}
\end{equation}
Assuming that in each two-body subsystem only one bound state exists, we can write the
asymptotic boundary conditions for the Faddeev component  $\phi$ as follows
\begin{equation}
\label{Asym}
        \phi(x,y,z)\sim \, \varphi_2(x)\,
        {\rm e}^{-k_y y}+A(x/y,z)\,
        \frac{{\rm e}^{-k_3(x^2+y^2)^{1/2}}}{(x^{2}+y^{2})^{1/4}}\ ,
\end{equation}
where  $\varphi _{2}(x)$ stands for the wave function of the two-body subsystem bound
state, $k_{y}=\sqrt{E_{2}-E_{3}}$, $k_{3}=\sqrt{-E_{3}}$,  $E_{2}$ is the two-body bound
state energy, and  $E_{3}$ the energy of the three-body  system. For three-body bound
states the first term corresponds to virtual decay into a particle and a two-body bound
system, while the second term corresponds to a virtual decay with an amplitude $A(x/y,z)$
into  three single particles. The term corresponding to the latter configuration can
generally be neglected for the states below the three-body threshold. Therefore, at
sufficiently large distances  $R_x$ and  $R_y$, the  asymptotic  boundary conditions for
the Faddeev component are
\begin{equation}
\label{ApprBC}
        \left .\frac{\partial }{\partial x}\ln\phi (x,y,z)
                \right |_{x=R_x}=-k_{x}\equiv i\sqrt{E_2}\,,\qquad
        \left .\frac{\partial }{\partial y}\ln\phi (x,y,z)
          \right |_{y=R_y}=-k_y\,.
\end{equation}
For bound state calculations Dirichlet or Neumann boundary conditions can also be employed.

The important property of the Faddeev components which makes them suitable for numerical
solution is their simple asymptotic form. For instance, each of the Faddeev components
holds only bound states of the corresponding two-body subsystem. In this respect the
${\bf x}_i$ coordinate is the internal coordinate of the corresponding two-body cluster
and the ${\bf y}_i$ coordinate plays the role of a reaction coordinate for all the states
below the 3-body (break-up) threshold. This simple physical meaning of the coordinates
suggest a natural requirement for discretizing the corresponding degrees of freedom: the
discrete analogs of the ${\bf x}_i$ coordinate should reproduce the spectrum of the
$i$-th cluster correctly, and discrete analogs of the ${\bf y}_i$ coordinate must
describe the scattering states reasonably well. These necessary requirements are easy to
check prior to performing actual calculations, and they also provide a solid ground for a
reasonable degree of automation for choosing the parameters of the numerical scheme.
Another advantage of the Faddeev equations is the asymptotic decoupling of the
components. The right-hand side of the equation~(\ref{eq:Faddeev}) is, roughly speaking,
exponentially small if the third particle is at larger distance than the typical size of
the two-body bound state. This means that at longer distances $|{\bf y}_i|>y_{\rm max}$
the Faddeev components rapidly decouple, and calculations can be performed in the regions
as small as the size of the largest two-body subsystem bound state.

The advantages of the Faddeev approach can be exploited even further when dealing with
short-range interactions; i.e. assuming that the potentials $V_i$ are zero (or
negligible) outside the region $|{\bf x}_i|<x_{\rm max}$. To clarify this, consider the
component $\Phi_i$ in the asymptotic region $|{\bf y}_i|\rightarrow \infty $, ${\bf x}_i
\in \sup V_i$, where it satisfies a Schr\"odinger equation with the corresponding channel
Hamiltonian
\[
  (H_0+V_i({\bf x}_i)-E) \Phi_i({\bf x}_i, {\bf y}_i) \approx 0 \; ,
\]
where $H_0$ is the free three-body Hamiltonian. This property of $\Phi_i$ suggests that,
rather than calculate $\Phi_i$ directly, we instead calculate $\tau_i$ (Eq. \ref{tauComp}),
which is better localized in configuration space
\begin{equation}
  \tau_i \equiv (H_0+V_i({\bf x}_i)-E) \Phi_i({\bf x}_i, {\bf y}_i) \; .
\label{tauComp}
\end{equation}
These $\tau_i$ are non-zero only for small $x_i\equiv|{\bf x}_i|<x_{\rm max}$ and small
$y_i\equiv|{\bf y}_i|<y_{\rm max}$. Accordingly, they are localized to a region that can
be {\em much smaller than the typical size of a two-body bound state}. This feature leads
to substantial computational savings~\cite{myCPL2}. The $\tau_i$ satisfy the following
integral equations
\begin{equation}
  \tau_i=-V_i \sum_{j\ne i} R_{2j}(E) \tau_j \; ,
  \label{eq:LCM}
\end{equation}
where $R_{2j}(E)$ are the resolvents of the corresponding channel Hamiltonians. Since
$\sup \tau_i \subset \sup V_i$, the $\tau_i$ are more suitable for numerical
approximation than the original Faddeev components. Furthermore, if the equation is being
solved using an iterative technique, then no explicit representation for the integral
operators $R_{2j}(E)$ is required. In this case we only need to calculate the action of
the integral operator on the $\tau_i$, which can be done with high computational efficiency
by numerically solving the corresponding differential equation with appropriate boundary conditions.
We call this computational scheme a Localized Component Method (LCM).

\section{Computer code}
In order to construct a discrete analogue of the system of equations (\ref{eq:LCM}) we
employ quintic Hermite splines together with the orthogonal collocations
method~\cite{deBoorSwartz}. A detailed description of the procedure is given in
\cite{RoudnevYak}. This high-order method guarantees fast convergence with respect to the
number of grid points, sparse matrix structure for the discrete analog of the
equation~(\ref{eq:Faddeev}), and fast calculation of matrix elements that makes it
possible to avoid storing big matrices in computer memory.

\begin{figure}
 \includegraphics[height=.4\textheight,clip=true]{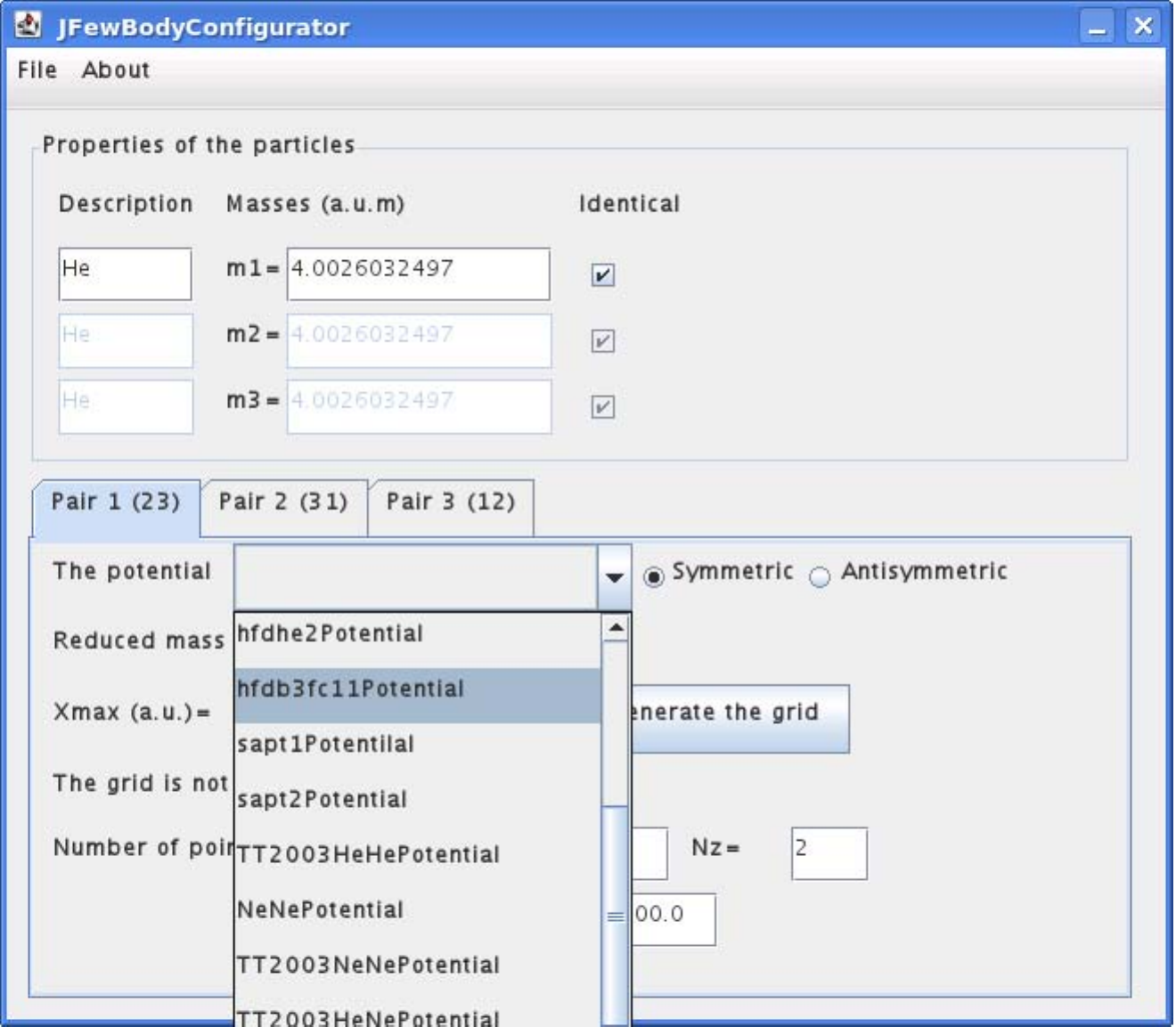}
 \caption{A screenshot of the configurator.\label{fig:Screenshot}}
\end{figure}

The code is written in Java and consists of two parts. The first part is a configurator
that simplifies composing the necessary configuration files. The configurator allows
the user to set masses of the interacting atoms, to specify identical particles in the system,
to choose a potential model, to set the cutoff distances $R_x$ and $R_y$ and to set the number of
grid points to be employed in the calculation.
It also generates a mesh with $L_2$-optimal point distribution which ensures the best possible
approximation of the three-body wave function in the asymptotic region.
In Fig.~\ref{fig:Screenshot} we show an example of the configurator screenshot.

The second part is the three-body computational kernel based on the LCM approach. The
kernel is currently capable of three-body calculations below the three-body threshold with a
limitation of no more than one two-body bound state contributing to each asymptotic
channel. This includes bound states, elastic scattering and chemical reactions below the
first vibrational excitation threshold. Typical computational time can take from minutes
to hours, depending on the physical system and the size of the grid.

\section{Results}

We apply the code to the calculation of binding energies of the Helium trimer $^4$He$_3$
three-atomic system, to verify that we can reproduce the known calculated properties of helium
trimer ground and excited states. This system is very particular about the approach being used, as
the trimer binding energy is extremely small and a large volume of the
configuration space should be treated, but the interaction features very strong repulsion at
short distances, which requires very precise numerical methods to be used.

Experimentally, helium dimers have been observed for the first time in 1993 by the
Minnesota group~\cite{DimerExp}, and in 1994 by Sch\"ollkopf and Toennies \cite{Science}.
Later on, Grisenti {\em et al.} \cite{exp} measured a bond length of $52 \pm 4$ {\AA} for
$^4$He$_2$, which indicates that this dimer is the largest known diatomic molecular
ground state. Based on this measurement they estimated a scattering length of
$104^{+8}_{-18}$ {\AA} and a dimer energy of $1.1^{+0.3}_{-0.2}$ mK \cite{exp}. In the
latter investigation~\cite{HeEfimovExp} the trimer pair distance is found to be
$1.1^{+0.4}_{-0.5}$ nm in agreement with theoretical predictions for the ground state.

Many theoretical calculations of these systems were performed for various interatomic
potentials \cite{Aziz91,TTY}. Variational \cite{Orlandini,Bressanini}, hyperspherical
\cite{EsryGreen,BlumeGreene,Barletta09} and Faddeev techniques
\cite{myCPL2,RoudnevYak,kea,SalciLevin,Lazauskas,MSSK} have been employed in this
context. It was found that the Helium trimer has two bound states of total angular
momentum zero: a ground state of about $126$ mK and an excited state of Efimov-type of
about $2.28$ mK. Experimentally this Efimov-type\cite{Efimov} excited state has not yet
been observed (see, e.g., \cite{Review2009} and refs. therein). It should be mentioned,
however, that the year 2006 is noticeable due the first convincing experimental evidence
for the Efimov effect in an ultracold gas of Caesium atoms \cite{Nature2006,Nature_th}.

In present calculations we employed the code based on the Faddeev differential
equations (\ref{H00}) with boundary conditions (\ref{ApprBC}). As He-He interaction we
used the semi-empirical LM2M2 potential~\cite{Aziz91}.  We use $m_{_{^4{\rm
He}}}=4.0026032197$~$a.u.m$ for the mass of the $^4$He atom and
$\frac{\hbar^2}{m_{_{^4{\rm He}}}k_B}=12.11928$\,K\,\AA$^2$, unlike many three-body
calculations, see, e.g.,~\cite{Review2009}, where a rounded value of the coefficient has
been used.

\begin{figure}
%\vspace*{-2cm}
\includegraphics[height=.4\textheight]{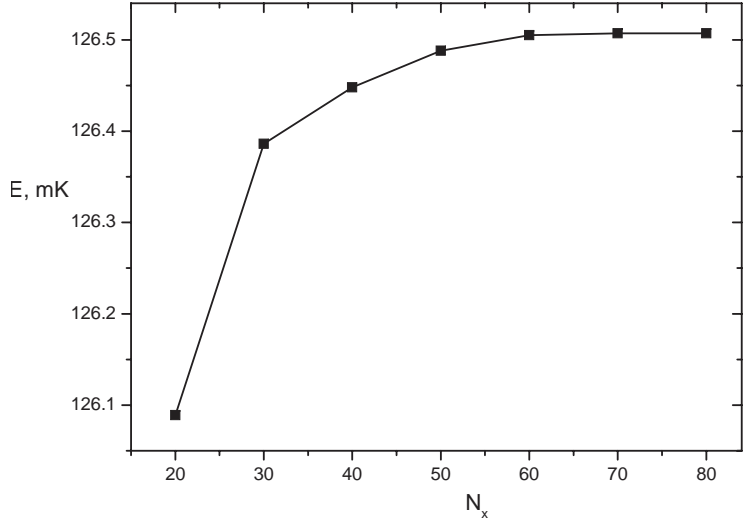}
\caption{Convergence of the Helium trimer ground state energy on the grids of $N_x = N_y$
points; $N_z = 5$.  \label{fig:converg}}
\end{figure}

Investigation of the bound state energy convergence with respect to the number of grid points
demonstrates that even a moderate number of points in variables $x$ and $y$ is sufficient to
get up to six accurate figures for the energy of the ground state (Fig.~\ref{fig:converg}).

%\begin{minipage}{7cm}
%\end{minipage}
%\begin{minipage{6cm}
%%%%%%%%%%%%%%%   TABLE I: DimerLen  %%%%%%%%%%%%%%%%%%%%%%%%%%%%%%
%\vspace*{-1.5cm}
\begin{table}
\begin{tabular}{ccccc}
\hline
 Potential model & $\varepsilon_d$ (mK) &  $\ell^{(1+1)}_{\rm sc}$ (\AA) & $<R>$ (\AA) & $\sqrt{<R^2>}$ (\AA) \\
\hline
LM2M2 \cite{Aziz91} & $-1.30348$ &  100.23& 52.001 & 70.926\\
\hline
Exp. \cite{exp} & $-1.1^{+0.3}_{-0.2}$  &$104^{+8}_{-18}$ &$52^{+4}_{-4}$ & -\\
 \hline
\end{tabular}
\caption{Dimer energy $\epsilon_d$,  $^4$He$-^4$He scattering length $\ell_{\rm
sc}^{(1+1)}$, bond length $<R>$ and root mean square radius $\sqrt{<R^2>}$ for the
potentials used, as compared to the experimental values of Ref. \cite{exp}.}
\label{tableDimerLen}
\end{table}

The $^4$He dimer binding energies, $^4$He--$^4$He scattering lengths and mean values of
the radius $<R>$ and $\sqrt{<R^2>}$ obtained with the LM2M2 potential~\cite{Aziz91}
are shown in Table \ref{tableDimerLen} in comparison with experimental data~\cite{exp}.
All the values agree with an experimental estimation of Ref.\cite{exp} within quoted
errors. The scattering length  $\ell^{(1+1)}_{\rm sc}$ of the system is bigger than the
range of the potential by an order of magnitude. All these features characterize the
Helium dimer as the weakest, as well as the biggest, diatomic molecule found so far. Due
to the fact that the energy of the dimer is so small, one should expect that the
$E_{^4{\rm He}_3}$ trimer indeed possesses the theoretically predicted state of the
Efimov type (see, \cite{Efimov,Review2009}).

In Table~\ref{tableTrimer} the results of trimer binding energies calculations obtained
with LM2M2 potential are summarized. The binding energies of the $^4$He trimer ground
($E_{^4{\rm He}_3}$) and exited ($E_{^4{\rm He}_3}^{*}$)  states are presented.
These results demonstrate good agreement between different methods and show that the
code competes well even against variational methods. It should be mentioned
that the energy estimates obtained with the code are non-variational, and further
variational improvements of the results are possible.

%\begin{table}%
%\begin{tabular}{cccccccccc}
%\\
%\hline \\[-2ex]
%   & present & \cite{kea} & \cite{RoudnevYak} & \cite{BlumeGreene}  &  \cite{Barletta09}  & \cite{SalciLevin} & \cite{Lazauskas} &\ \cite{Orlandini}  \\[0.1ex]
%\hline \\[-2.5ex]
%$\bigl|E_{^4{\rm He}_3}\bigr|$ (mK) &\  $126.507$ &\ $126.45$  &\ 126.41 &\
%$125.52$\tablenote{In original paper the energy value is given in cm$^{-1}$ (1cm$^{-1}$ =
%1.4387752 K).}   &\   126.15  & $126.2$ &\ 126.39 &\ $125.6^\mathrm{*}$   \\[1ex]
%$\bigl|E_{^4{\rm He}_3}^{*}\bigr|$ (mK) &\  $2.276$ &\ 2.282\tablenote{This value was
%rounded in \cite{MSSK}.}  &\  2.271&\    &\  2.274 &  & 2.268 &\ $2.245^\mathrm{*}$\\[1ex]
%\hline
%\end{tabular}
%\caption {Results for binding energies of the $^4\mathrm{He}_3$ trimer for LM2M2
%potential.} \label{tableTrimer}
%\end{table}

\begin{table}
\begin{tabular}{ccccccccccc}
\\
\hline \\[-2ex]
   & present & \cite{kea} & \cite{RoudnevYak} & \cite{SalciLevin}  &\cite{Lazauskas} & \cite{MSSK}
   & \cite{BlumeGreene}  &  \cite{Barletta09}    &\ \cite{Orlandini}  \\[0.1ex]
\hline \\[-2.5ex]
$\bigl|E_{^4{\rm He}_3}\bigr|$ (mK) &\  $126.507$ &\ $126.45$  &\ 126.41 &\  $126.2$ &\
 126.39 &\ 125.9 &\ $125.52$\tablenote{In original paper the energy value is given in
cm$^{-1}$ (1cm$^{-1}$ = 1.4387752 K).}  &\   126.15    &\ $125.6^\mathrm{*}$   \\[1ex]
$\bigl|E_{^4{\rm He}_3}^{*}\bigr|$ (mK) &\  $2.276$ &\   &\  2.271&\
 &\ 2.268  & 2.282& &\  2.274    &\ $2.245^\mathrm{*}$\\[1ex]
\hline
\end{tabular}
\caption {Results for binding energies of the $^4\mathrm{He}_3$ trimer for LM2M2
potential.} \label{tableTrimer}
\end{table}

We are planning to continue testing the code within current applicability limits,
including scattering calculations, systems of distinguishable particles and modeling
clusters of other rare gas atoms.

%%%%%%%%%%%%%%%%%%%%%%%%%%%%%%%%%%%%%%%%%%%%%%%%
%% BACKMATTER
%%%%%%%%%%%%%%%%%%%%%%%%%%%%%%%%%%%%%%%%%%%%%%%%
\begin{theacknowledgments}
This work is supported by Heisenberg-Landau Program (EK) and the NSF grant PHY-0903956
(VR and MC).
\end{theacknowledgments}



\bibliographystyle{aipproc}   % if natbib is available

%\bibliography{sample}

%  \typeout{******************************************}
%  \typeout{** Please run "bibtex \jobname" to optain}
%  \typeout{** the bibliography and then re-run LaTeX}
%  \typeout{** twice to fix the references!}
%  \typeout{******************************************}

%\end{document}

%%%%%%%%%%%%%%%%%%%%%%%%%%%%%%%%%%%%%%%%%%%
%% The following lines show an example how to produce a bibliography
%% without the help of the BibTeX program. This could be used instead
%% of the above.
%%%%%%%%%%%%%%%%%%%%%%%%%%%%%%%%%%%%%%%%%%%

\end{document}